\documentclass[twocolumn,preprintnumbers,amsmath,amssymb,superscriptaddress,prl]{revtex4}

\newcommand{\eps}{\varepsilon}
\newcommand{\om}{\omega}
\newcommand{\ka}{\kappa}
\newcommand{\be}{\beta}
\newcommand{\del}{\delta}
\newcommand{\epso}{\eps_0}
\newcommand{\omo}{\om_0}
\newcommand{\delo}{\del_0}
\newcommand{\epspar}{\eps_{\|}}
\newcommand{\epsper}{\eps_{\bot}}
\newcommand{\kp}{k^{\prime}}
\newcommand{\kpp}{k^{\prime\prime}}

\usepackage{dcolumn}% Align table columns on decimal point
\usepackage{bm}% bold math
\usepackage[latin1]{inputenc}
\usepackage[dvips]{graphicx}
\usepackage{float}
\usepackage[dvips]{epsfig}
\usepackage[latin1]{inputenc}

\begin{document}

\title{Coherent emission from a disordered organic semiconductor induced by
    strong coupling with surface plasmons}
\author{S. Aberra Guebrou}
\author{C. Symonds}
\author{E. Homeyer}
\author{J.C. Plenet}
\affiliation{LPMCN; Université de Lyon; Université Lyon 1 and CNRS, UMR 5586; F-69622 Villeurbanne, France}
\author{{Yu}.N. Gartstein}
\affiliation{Physics Department, University of Texas at Dallas, Richardson, Texas 75080, USA}
\author{V.M. Agranovich}
\affiliation{Chemistry Department, University of Texas at Dallas, Richardson, Texas 75080, USA}
\affiliation{Institute of Spectroscopy, Russian Academy of Sciences, Troitsk, Moscow Region, 142190, Russia}
\author{J. Bellessa}
\email{joel.bellessa@univ-lyon1.fr} \affiliation{LPMCN; Université
de Lyon; Université Lyon 1 and CNRS, UMR 5586; F-69622
Villeurbanne, France}

\date{\today}

\begin{abstract}

In this Letter we show that the strong coupling between a disordered set
of molecular emitters and surface plasmons leads to the formation
of spatially coherent hybrid states extended on macroscopic distances. Young type interferometric experiments performed on a system of
J-aggregated dyes spread on a silver layer evidence the coherent
emission from different molecular emitters separated by several
microns. The coherence is absent in systems
in the weak coupling regime demonstrating the key role of the
hybridization of the molecules with the plasmon.

\end{abstract}

%Use showkeys class option if keyword

\maketitle
Localized and delocalized surface plasmon (SP) modes feature stronger confined electric fields and can therefore efficiently
interact with molecules or semiconductors close to metallic interfaces \cite{LukinNat,WogPRL99}. This effect has been widely used to modify optical properties of different types of emitters \cite{NovotnyPRL} resulting in the radiative rate enhancement and strong coupling. Another important but less studied aspect is the interaction between different emitters induced by SPs. Efficient energy transfer between donor and acceptor molecules  on
opposite sides of metal films has been demonstrated \cite{BarnesScience}, the transfer being mediated by symmetric and antisymmetric SP modes on a metallic film. A cooperative emission, similar to Dicke superradiance, of an ensemble of dipoles in the vicinity of a metallic nanosphere has also been theoretically identified \cite{Pustovit}.

The strong coupling regime offers an interesting perspective for molecular coupling \cite{LidzeyScience} and collective emission.
This regime has been reported in various disordered materials such as aggregated dyes coupled to a cavity photon \cite{LidzeyPRL99} and organic materials \cite{BellessaPRL,DintingerPRB,BaumbergPRL,HakalaR6G} and semiconductor quantum dots \cite{GomezCdSe} interacting with SPs. Despite the difference of electronic properties of the emitting species in these systems, they all can be seen as a set of localized (no well-defined wavevector) emitters coupled to a propagating extended mode. The strong interaction between a disordered medium and an electromagnetic wave in microcavities has been shown theoretically \cite{AgranovichPRB67} to result in hybrid eigenstates that are quantum superpositions of a photon and excitations on a large number of molecular sites. Experimentally, the strong coupling is usually identified via observations of an anticrossing in the dispersion relation. The dependence of the Rabi splitting on the concentration of emitters is a good indication of collective effects, however the  coherence of spatially remote emitters induced by hybridization with a  plasmon (or cavity) mode has never been directly evidenced. In fact, as was theoretically illustrated \cite{AgranovichPRB75,Litinskaya}, disorder may result in relatively small modifications of the energetic spectrum of hybrid states, polaritons, while drastically affecting the extent of the polariton wavefunction.

In this Letter we specifically address the issue of spatial coherence and find experimental evidence that the strong interaction between a set of molecular emitters and a SP can lead to the formation of a macroscopic extended coherent state akin to that in a macroscopic polymer chain \cite{DubinNatPhys}. We investigate the diffusion and the spatial coherence of the emission of J-aggregated dyes on silver with Young-type interferometric experiments, evidencing an in-phase emission of localized emitters separated by several microns. The extension of such coherent states over a large number of molecules can conceivably be of significance for long-range energy transfer as well as for new optical alloys where the properties of different kinds of emitters spatially separated on a metal film would be mixed via the plasmon.

Our experiments were performed using a leakage radiation (LR)
microscopy setup \cite{Hecht}, presented in Fig.~1(a).
The samples are excited with a laser
light at 532 nm focused on the top side of the sample. The LR of the emission
through the silver layer is collected with an immersion oil
microscope objective (NA=1.49) placed in contact with the bottom
side of the sample, and imaged either on a CCD camera or on a CCD
detector associated with a spectrometer. A notch filter at 532 nm suppresses the laser radiation in the images. The LR
microscopy technique allows a Fourier space imaging providing the
angular dispersion of the radiated emission \cite{Drezet} as well
as a direct imaging of the emission. In this latter case, a beam
block can be inserted in the Fourier plane of the collection
optics in order to filter the radiated wavevectors, enabling a
selective imaging of the SPs propagating in a given direction.
During this study, the experiments were conducted by collecting
either all the wavevector components of LR (k$_{total}$
configuration), or solely the wavevector components corresponding
to a propagation in the upper half-space as shown in the inset of
Fig.~1(a) (k$_{up}$ configuration).

% The excitation spot has a FWHM  of 0.7 $\mu$m.

\begin{figure} \includegraphics[width=7.5cm]{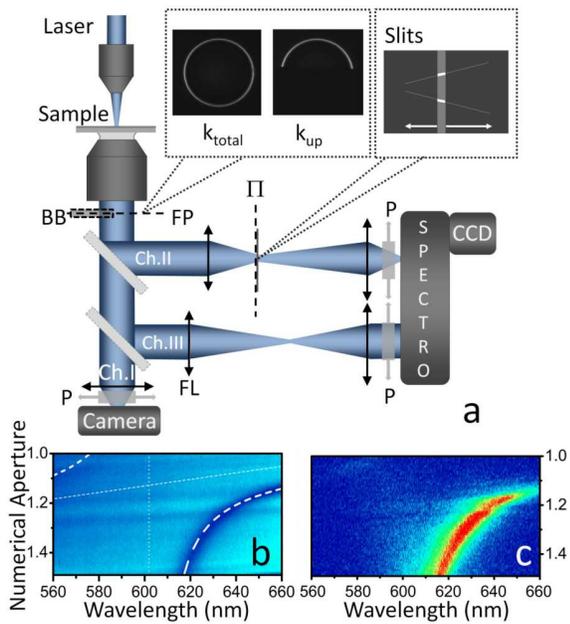}
\caption{  (a) Experimental setup. A
direct imaging is performed on the channel I. The channel II is
devoted to coherence measurements by inserting slits in an
intermediate image plane $\Pi$ of the sample surface. The channel III is used for dispersion measurements, by imaging the Fourier plane of the microscope objective on the entrance slit of the spectrometer. A beam block
(BB) can be inserted in the Fourier plane (FP) to select a given
wavevector direction for all the channels. (b)
Reflectometry and (c) Luminescence spectra of a TDBC dye layer on a
silver film (sample C), displayed as a function of the wavelength
and of the numerical aperture (channel III). The doted lines correspond to the bare plasmon and exciton wavelengths, and the dashed line to the calculated polaritonic dispersion.}
\end{figure}

Coherence of the electric field pattern on metal surfaces is associated
with highly-delocalized surface modes. This
coherence is related to the spatial propagation of the SP but not
to the coupling as such between different emitters. This phenomenon has
been observed in the infrared range \cite{Carminati} for thermal
emitters but can also be present in the optical range \cite{Brongbergsma}. In order to
clearly evidence the effect of the SP-exciton hybridization,
samples in the weak and strong coupling regimes have been compared.

\begin{figure} \includegraphics[width=7.0cm,height=4.0
cm]{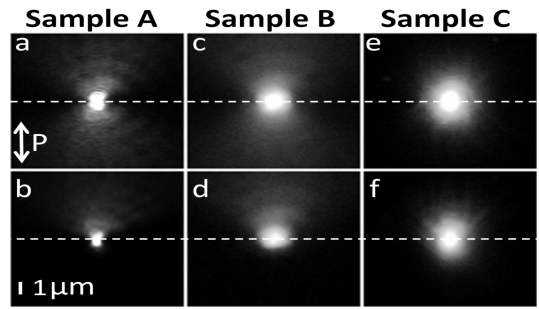}
\caption{Direct images of the leakage radiation on
the surface for the different samples: (a) and (b) sample A (insulated microspheres). (c)
and (d) sample B (CdSe nanocristals). (e) and (f), sample C (TDBC layer
in strong coupling). Panels (a), (c) and (e) correspond to the $k_{total}$
detection and panels (b),(d) and (f) to $k_{up}$ detection
configuration. The vertical arrow indicates the polarization
direction P. The horizontal dotted lines correspond to the
position of the emission center.}
\end{figure}

Three samples were investigated, each elaborated by spin coating
an active layer onto a 45 nm silver film thermally evaporated on a
glass coverslip. The active layer of the first sample (sample A)
is composed by widely separated fluorescent polystyrene
microspheres, emitting at 560 nm and having a diameter of 100 nm.
The distance between each microsphere is sufficient to consider
the emission of an insulated nanoparticle. The second sample
(sample B) contains a continuous layer of CdSe quantum dots
emitting around 660 nm. The distance between the quantum dots is
smaller than the resolution of our setup, so that an uniform
emission pattern will appear on the images. From the dispersion
relations obtained by reflectometry measurements on these two
samples, it has been checked that both kind of emitters are in
weak coupling regime with the SP (data not shown) \cite{CdSe}. In the third
sample (sample C) the optically active component is a J-aggregated
$5,5^{'} ,6,6^{'}
-tetrachloro-1,1^{'}-diethyl-3,3^{'}-di(4-sulfobutyl)-benzimidazolo-carbocyanine$
(TDBC). The TDBC layer is formed by an ensemble of linear
J-aggregates chains with a small length compared to the
wavelength, and thus can be seen as a set of independent emitters
randomly spread throughout the film. Figure 1(b)
presents the reflectometry experiment recorded in the Fourier
space for sample C. The measured dispersion relation presents an
anticrossing characteristic of a strong coupling regime between
the SP and the TDBC exciton, as previously reported for this kind
of sample\cite{BellessaPRL}. The Rabi splitting energy calculated with a
classical two level oscillator model is 300 meV.

In a first set of experiments the emission through the silver
layer for each sample was imaged on the CCD camera. For the sample
A in the k$_{total}$ configuration (Fig.~2(a)), the emission of one microsphere
forms two lobes oriented in the vertical direction. When the beam block is
inserted in the Fourier plane (k$_{up}$), only the radiation
having wavevector components in the upper half-plane is detected.
In this case the lower emission lobe is suppressed (Fig.~2(b)).
These images can be clearly interpreted in terms of a localized
particle emitting SPs propagating upwards and downwards (Fig.~2(a))
or only upwards (Fig.~2(b)). The same behavior is observed for
sample B (Fig.~2(c),(d)). In these images the bright central area
corresponds to a collection of CdSe quantum dots excited by the
laser beam. The emission pattern presents two lobes, the lower one
being suppressed in k$_{up}$ configuration (Fig.~2(d)). The set of
emitters in the weak-coupling regime with SP modes behaves as a
sum of independently emitting particles. The recorded emission
patterns are drastically different when the emitters are in the
strong-coupling regime with SPs, exhibited by sample C (Fig. 2
(e),(f)). In this case, the emission pattern has an oval shape in the
k$_{total}$ configuration and is only slightly modified in the
k$_{up}$ configuration. The persistence of the lower lobe in the
k$_{up}$ configuration indicates that the emission can not be
interpreted in terms of a set of particles which independently
emit propagating SPs but rather as arising from an ensemble of
emitters all coupled to the same SP resulting in extended
hybridized polaritonic states.

In order to study the spatial coherence of this emitter assembly,
two Young's slits are inserted in an intermediate image plane of
the sample. The Youg slits are obtained by crossing a vertical slit with two V shaped slits in order to vary the interslit distance. The resulting slits, located on both sides of the excitation spot (full width at half maximum FWHM 0.7 $\mu$m), select the
emission from two regions of the sample separated by a distance of
2.8 $\mu$m. The interference pattern recorded from sample A (a
single localized emitter) is shown in Fig.~3(a). Interference
fringes appear on the image. This case is analogous to a classical
wavefront division interference experiment: the emitter generates
SPs propagating upwards and downwards before reaching the spatial
regions selected by the upper and lower slits. The radiation
propagating along these two paths interferes on the entrance slit
of the spectrometer. In the k$_{up}$ configuration, however, which
disables the lower direction of propagation, only the upper slit
remains illuminated and the interference fringes disappear
(Fig.~3(b)). In the case of sample B no fringes are visible in both
configurations. This is what is expected for a set of independent
localized emitters. Each of them generates its own interference pattern associated with a phase difference wich depends on its position between the slits. As the excitation spot extends over 0.7$\mu$m, a blurring of the fringes occurs.

In contrast to samples A and B, sample C displays interference
patterns in both k$_{total}$ (Fig.~3(e)) and k$_{up}$ (Fig.~3(f))
configurations between 610 nm and 645 nm, corresponding to
 the SP-exciton mixed (polaritonic) states.
The persistence of the interference fringes, even when the
SP-assisted interferences are suppressed (k$_{up}$), demonstrates
the spatial coherence of the extended hybrid states over a
distance of 2.8 $\mu$m. The results may be compared to similar
experiments performed on long polymer chains \cite{DubinNatPhys}.
For those long molecules the emission along the chain is coherent:
the whole polymer chain emits even when excited with a localized
spot. In a comparable manner, in our case a single dye chain can
not be excited independently from the others as shown in the
diffusion experiments of Fig.~2.

\begin{figure} \includegraphics[width=8.0cm]{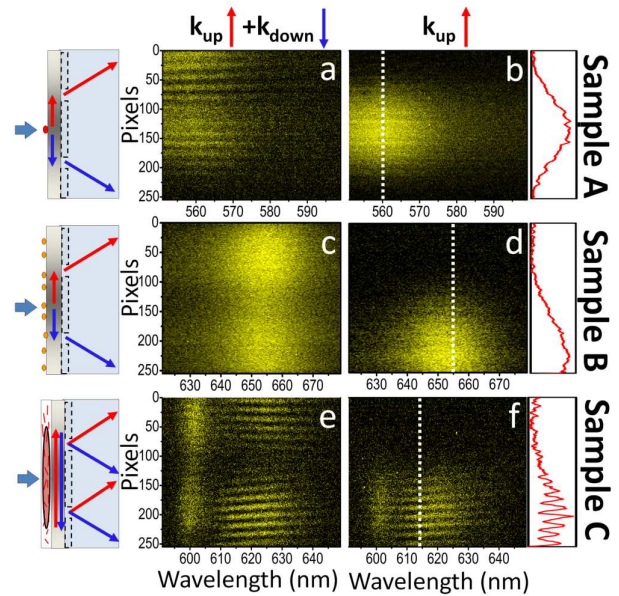}
\caption{Interference pattern recorded for
samples A, B and C: without selection on the wavevector (panels (a),
(c) and (e), respectively) and with only the upwards propagation
(panels (b), (d) and (f), respectively). Layouts of different
propagation mechanisms are drawn on the left side of the figure
for each of the samples. For the (k$_{up}$) configuration, the intensity profile measured along the white dotted line is drawn on the right side of the figure.}
\end{figure}

The spatial extent of the SP-exciton hybrid states is now further
addressed with a similar Young experiment but with a laser spot
(FWHM 10 $\mu$m) covering both interfering regions on the sample.
The emission pattern of a control sample consisting of a TDBC
layer directly deposited on a glass cover slip without silver is
shown in Fig.~4(a) and does not display any interference fringes,
which agrees with the assertion that the TDBC layer is constituted
by independent emitters formed by the aggregated dye chains.
Interference images are recorded for sample C, TDBC layer on
silver (Fig.~4(b)). Two different regions can be seen in the image:
the polaritonic emission between 610 and 645 nm presenting interference
fringes and the emission around 600 nm without interferences. The
latter emission comes from incoherent states at the bare TDBC
exciton energy. The existence of both incoherent and coherent
extended states in the polariton spectrum has been discussed
theoretically \cite{AgranovichPRB67,Litinskaya,AgranovichPRB75}
for organic microcavities in the strong-coupling regime. Fig.~4(c)
presents the visibility of the fringes as a function of the
distance between the interfering regions. For a detection
wavelength of 630 nm, the visibility vanishes at 7.5 $\mu$m, while
for 610 nm the visibility decreases faster and vanishes around 6.5
$\mu$m showing a reduction of the coherence length when the
wavelength becomes closer to the bare exciton emission at 600 nm.

\begin{figure} \includegraphics[width=8 cm,height=2.8
cm]{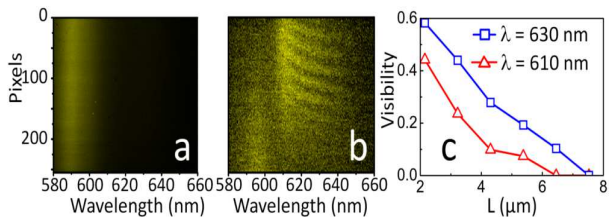}
\caption{(a) Emission pattern
for a TDBC layer directly deposited on a glass substrate. (b)
Typical interference pattern for a TDBC layer on silver with an
excitation spot larger than the interslit distance (here 4.4 $\mu$m).
(c) Visibility of the fringes as a function of the distance for two
detection wavelengths (610 and 630 nm).}
\end{figure}

To elucidate the generic character and the origin of the
frequency-dependent degree of coherence, we now provide an
illustrative calculation of surface polaritonic states in a
simpler parent system consisting of a thin (thickness $d$)
resonating organic layer  on a metallic substrate \cite{AgrMal74}.
The dispersion equation (DE), frequency $\om$ vs in-plane wave
vector $k$, can be found in this case  as
$$\ka_1/\eps_1 +
\ka_2/\eps_2 = \be \left(k^2/\epsper - \om^2/c^2 + \epspar
\ka_1\ka_2/\eps_1\eps_2 \right),$$
where $\ka_i^2=k^2 - \eps_i
\om^2/c^2$, $\be = -\tanh(\ka d)/\ka$ and
$\ka^2=(\epspar/\epsper)k^2-\epspar \om^2/c^2$. The DE features various
$\om$-dependent dielectric functions, their imaginary
parts  being representative of the
dissipation and disorder in the system. In the absence of the layer,
$\be = 0$, the DE would yield standard SP excitations at the
interface between dielectric (we choose vacuum $\eps_1=1$) and
metallic ($\eps_2 = \eps_{2b} - \om_p^2/\om(\om + i\del)$)
half-spaces. The organic layer here is anisotropic: for layered
J-aggregates we may assume that the resonance (frequency $\omo$)
occurs only for the the polarization along the layer
($\epspar=\epso + A/\left(\omo^2 - (\om+i\delo)^2 \right)$), while
no resonance polarization takes place perpendicular to the layer
($\epsper=\epso$).

%To elucidate the generic character and the origin of the
%frequency-dependent degree of coherence, we now provide an
%illustrative calculation of surface polaritonic states in a
%simpler parent system consisting of a thin resonating organic
%layer on a metallic substrate \cite{AgrMal74}. The dispersion
%equation relating polariton frequency $\om$ to its in-plane
%wavevector $k$  features various $\om$-dependent dielectric
%functions $\eps$ of the system components: they are $\eps_1=1$ of
%vacuum, $\eps_2 = \eps_{2b} - \om_p^2/\om(\om + i\del)$ of the
%metallic substrate; for layered J-aggregates we may assume that
%the resonance (frequency $\omo$) occurs only for the the
%polarization along the layer ($\epspar=\epso + A/\left(\omo^2 -
%(\om+i\delo)^2 \right)$), while no resonance polarization takes
%place perpendicular to the layer ($\epsper=\epso$). The imaginary
%parts of the dielectric functions are representative of the
%dissipation and disorder in the system.

\begin{figure}
\includegraphics[width=7.5cm]{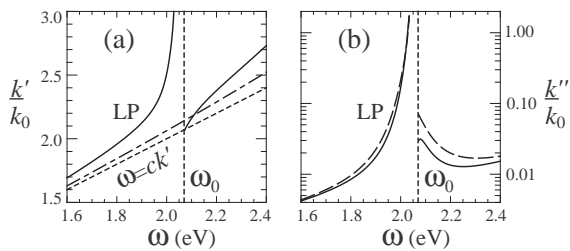}
\caption{Dispersion of surface
polaritons in the vicinity of the resonance with the excitations
of the thin layer: (a) Real and (b) Imaginary parts of the wave
vector as a function of frequency in units of $k_0$, the wave
number of the vacuum photon at 1 eV ($\lambda_0 \simeq 1240$ nm).
The Rabi splitting of two polariton branches is clearly seen in
panel (a), which also displays the bare (in the absence of the
layer) SPs by a dash-dotted line. Numerical parameters used:
$\omo=2.07$ eV, $\epso=3.05$, $d=20$ nm, $A=2$ eV$^2$, $\delo=14$
meV, $\eps_{2b}=3.29$, $\om_p=8.93$ eV, $\del=79$ meV. Dashed
lines in panel (b) show the results for larger $\delo=25$ meV.}
\end{figure}

Fig.~5 shows the resulting two polaritonic branches in the form of
the real and imaginary parts of the wavevector ($k=\kp+i\kpp$) as
functions of frequency $\om$ in the vicinity of the resonance
$\omo$. Displayed are the states beyond the light line ($\om <
c\kp$), which would be stationary states ($\kpp=0$) in the absence
of the dissipation in the metal ($\del=0$) and the layer
($\delo=0$). It is the non-vanishing dissipation (disorder) that
makes these states decaying, the decay length ($1/\kpp$)
establishing the spatial scale of the coherence. We attribute the
experimentally observed features to the behavior exhibited by the
lower-energy polaritons (LP) in Fig.~5 where the decay length can
indeed be in the $\mu$m range below the resonance frequency
(e.~g., $\sim 2.2$ $\mu$m for $\lambda=625$ nm in this example)
but strongly decreases upon the approach to the resonance. This
trend agrees with observations in Fig.~4.

The formation of macroscopic coherent states is expected for a
large number of disordered materials like organic dyes
\cite{BellessaPRL,HakalaR6G}, layers of quantum dots
\cite{GomezCdSe}, and rare earth ions \cite{Lipson} strongly
coupled to SPs. The simple design of the samples enables the
manipulation of these coherent states by structuring the active
layer at distances shorter than the coherence length, using
patterns or inclusion of several different materials, opening a
way to a new class of plasmonic materials.

The authors thank J. Bloch, R. Grousson and M.
Broyer for helpful discussions and support. This work has been
supported by the French ANR PNANO SCOP and by the Lyon Nanoptec
center. V.M.A. acknowledges hospitality of Scuola Normale Superiore (Pisa) via
support of the European Commission (Grant N. FP7-PEOPLE-ITN-2008-237900 ``ICARUS'')
and thanks Russian Foundation for Basic Research for partial support.


\begin{thebibliography}{}
\bibitem{CdSe} It has to be noted that the CdSe quantum dots can switch from weak to strong coupling if their concentration is increased by five orders of magnitude \cite{GomezCdSe}. Nevertheless another material, namely the TDBC, has been chosen for experiments in strong coupling. Indeed, the TDBC allows a Rabi splitting more than 4 times larger than the polariton linewidths, thus ensuring an interaction with plasmon largely prevailing the scattering and damping of the system.
\end{thebibliography}
\end{document}